\documentstyle[pra,preprint,epsfig,aps]{revtex}
\tighten
\begin{document}
\draft
\title{Operational Theory of the Eight-Port Homodyne Detection}
\author{Piotr Kocha\'nski}
\address{Center for Theoretical Physics,
Al. Lotnik\'ow 32/46, 02-668 Warsaw, Poland}
\author{Krzysztof W\'odkiewicz}
\address{Center for Advanced Studies and Department of Physics and
Astronomy, \\
University of New Mexico, Albuquerque, New Mexico 87131, USA \\
and \\
Instytut Fizyki Teoretycznej, Uniwersytet Warszawski,
Warszawa 00-681, Ho\.za 69, Poland \cite{uw}}
\maketitle
\begin{abstract}
The eight-port homodyne detection apparatus is analyzed in the framework of
the operational theory of quantum measurement.  For an arbitrary quantum
noise leaking through the unused port of the beam splitter,  the positive
operator valued measure and the corresponding operational homodyne
observables are derived. It is shown that such an eight-port homodyne device
can be used to construct the operational quantum trigonometry of an optical
field. The quantum trigonometry and the corresponding phase space Wigner
functions are derived for a signal field probed by a classical local
oscillator and a squeezed vacuum in the unused port.
\end{abstract}

\section{Introduction}
In this paper we present a general operational description of  measurements
performed with the help of the eight--port homodyne detector. This
experimental setup has been extensively
used in the past years in connection with the homodyne detection of optical
signals \cite{walker87} and the determination  of the phase of quantum states
of light\cite{nfm}. The series of beautiful experiments, performed by
Noh, Foug\`eres and Mandel (NFM) has provided a fully  operational approach
to the measurement of the quantum phase of a single mode light field. These
experiments have inspired a lot of further theoretical works devoted to a
general operational approach to   quantum measurements in quantum optics.

The NFM experiments have been  treated theoretically in many
different ways \cite{mf+ws93,mf+kv+ws93,ul+hp93,pr+kw94}, however
our approach will be entirely based on  the concept of  the operational
phase space {\em propensity} \cite{kw84} and on the associated with it
operational quantum observables \cite{bge+kw95}.
Such a formulation, which is one of many possible descriptions of  realistic
measurements, is closely connected with the concept of the positive operator
valued measure (POVM). In the operational formulation the POVM  replaces the
 spectral measure of the intrinsic quantum observable \cite{blm,peres}.

The operational formalism has  been applied to the
NFM experimental device in the context of  the operational definition of the
Hermitian phase operator \cite{bge+kw95}. In the framework of such an
approach, the so called quantum trigonometries for the optical fields have
been constructed \cite{bge+kw+pr95,pk+kw96,kw+bge96}. Recently the
operational theory  has been  applied to the homodyne
detection scheme with a fixed or a random phase of the local classical
oscillator\cite{kb+kw96}.

The advantage of using the operational measurement theory is that
we can easily and naturally take into account the influence of
the experimental device on the measured system. For example,
the eight port homodyne detector is characterized by the efficiency
of the detectors, the reflectance and the transmittance
of the beam splitters, and the state of the quantum field entering the unused
ports of the  device. This additional ``noise'' field plays
a fundamental role, if  the NFM device is used to investigate the phase of
quantum optical fields. In such a setup the impact of the correlations
on the probe field and the noise field
plays a crucial role in
the operational definition of the quantum phase.
This particular
interplay between the quantum signal and the quantum noise, leaking through
the unused port in the NFM device, provides the framework of the operational
approach to the quantum phase.
The case of the noise field being
the vacuum state has been analyzed, in the spirit of the operational
formalism, in Ref.\cite{bge+kw+pr95}.
These results have been generalized for the case of the
squeezed vacuum state \cite{pk+kw96}. In this paper  we give a more detailed
discussion of the squeezed vacuum  state case and present general results
obtained for an arbitrary  noise field.

In Section II we shortly review the operational approach to the
measurement in quantum mechanics. In Section III  we derive the operational
probability density ---  propensity for an arbitrary noise field in the NFM
apparatus. In Section IV we derive the operational homodyne observables for
the eight-port homodyne detection scheme. Section V is devoted to the
discussion of the quantum trigonometry, generated in the operational phase
measurements in the NFM device, with the noise being a squeezed vacuum.
Finally, some concluding remarks are presented.

\section{Operational Approach to Measurements in Quantum
Mechanics}
The approach to the quantum mechanical theory of measurement that
will be used in this paper was described in
details in Ref. \cite{bge+kw95}.
Below we just recall the main
results of this work in order to make the paper self-contained.
The statistical outcomes of an ideal
measurement of a certain observable $\hat{A}$: $\hat{A}|a\rangle=a|a\rangle$
are described by the spectral measure \cite{vneumann32}:
\begin{equation}
\label{intrinsic_probability}
p_{\psi}(a)=|\langle a|\psi\rangle|^{2}\,,
\end{equation}
where $|\psi\rangle\in {\cal H}$ is the state vector of the measured system.
It is known that the spectral measure contains all the relevant statistical
information about the investigated system but it makes no reference to the
apparatus employed in the actual measurement. Due to this property $\hat{A}$
will be called an {\em intrinsic quantum observable}.

The description of  more  realistic measurements necessarily involves
additional degrees of freedom such as, for example, the unused ports of the
beam splitters or the local oscillator being the phase   calibrating device.
In such an approach, in order to describe the quantum mechanical
measurement, we have to work in the {\em extended} Hilbert space
${\cal H}\otimes{\cal H}_{{\cal F}}$, containing both the Hilbert
space of the investigated system $\cal{H}$ and the Hilbert space  ${\cal
H}_{{\cal F}}$ of all the additional measuring devices, which will
be called for short a {\em filter}.
The process of measurement is described by the interaction between the system
and the filter. Because we are interested only in the properties of the
system, we shall reduce the  degrees of freedom by tracing the
dynamical evolution in the combined Hilbert space over the filter Hilbert
space obtaining in such a way an operational {\em propensity} $\Pr(a)$, which
is nothing else but a probability distribution of a certain classical
quantity $a$ measured in a real experiment. This propensity, for arbitrary
density operator $\hat{\rho}$ of the system, is
\begin{equation}
\label{propensity}
\Pr(a)={\rm Tr}\{\hat{{\cal F}}(a)\hat{\rho}\},
\end{equation}
with the normalization condition
\begin{equation}
\int da\,\Pr(a)=1 \Longleftrightarrow\int da\,\hat{{\cal F}}(a)=\hat{1},
\end{equation}
where $\hat{{\cal F}}(a)$ is a filter dependent POVM.
We see that in a realistic measurement, involving a filter, the spectral
decomposition
of $\hat{A}$ is effectively replaced by a positive valued measure
$da\,\hat{{\cal F}}(a)$ \cite{blm}.

In view of the linear relation between the propensity and the POVM, the
operational statistical moments of the measured quantity
\begin{equation}
\label{moments}
\overline{a^{n}}=\int da\,a^{n}\Pr(a)\,,
\end{equation}
define uniquely an algebra of {\em operational operators}
\begin{equation}
\label{opoperator}
\overline{a^{n}}\equiv \langle \hat{A}^{(n)}_{{\cal F}}\rangle,
\end{equation}
where
\begin{equation}
\label{op_operator}
\hat{A}^{(n)}_{{\cal F}}=\int da\,a^{n}\hat{{\cal F}}(a).
\end{equation}
These operators are called  operational operators, because they
represent quantities measured in a real experiment i.e., described by a
dynamical coupling of the measured system and the filter.
As a rule, the algebraic properties of the set $\hat{A}^{(n)}_{{\cal F}}$
differ significantly from those of the powers of $\hat{A}$.
For instance,
$(\hat{A}^{(1)}_{{\cal F}})^{2}\neq\hat{A}^{(2)}_{{\cal F}}$. This property
will have important consequences in the discussion of the uncertainty
introduced by the measurement.

It is seen that the propensity and the operational operators
are natural equivalents of the spectral probability distribution and the
intrinsic operators. The difference being that the operational observables
carry information  about the system under investigation and the selected
measuring device, which is represented in this paper by the
eight-port homodyne detector.

\section{NFM Apparatus with an Arbitrary Noise Field}

In this Section we derive a general formula for the POVM that describe the
NFM device with an arbitrary noise filed leaking through the unused port of
the beam splitter. We shall assume that the NFM apparatus has unit detectors
efficiency, that a 50\%:50\%  beam splitters are used and that the local
oscillator is classical.
We assume that we want to measure an observable
$\hat{A}(\hat{b},\hat{b}^{\dagger})$. In general such an observable requires
additional informations about the particular ordering of the boson creation
$\hat{b}^{\dagger}$ and  annihilation $\hat{b}$ operators. For example the
quantum observable that corresponds to the polar decomposition of the
amplitude of a field mode could be defined as ${\hat{b}}/
\sqrt{\hat{b}^{\dagger}\hat{b}}$, however this formula is not unique because
of the possible different ordering of the boson operators.
Due to the ordering problem,
there is no canonical way of selecting the right, i.e., the physically
justified ordering. The operational approach removes this ordering ambiguity.
We shall perform a measurement of this observable using the NMF device.
In this case the filter Hilbert space is reduced to a Hilbert space
${\cal{H}}_{v}$, where $v$ denotes the quantized mode leaking through the
unused port.
In the  extended filter--system space ${\cal{H}}_{b}\otimes{\cal{H}}_{v}$ the
observable $\hat A$ will be the following function of the creation
and annihilation operators
\begin{equation}
\label{o}
\hat{A}=\hat{A}(\hat{b}+\hat{v}^{\dagger},\hat{b}^{\dagger}+\hat{v}),
\end{equation}
where the operators $\hat{b},\hat{b}^{\dagger}$ represent the probe field
and the operators $\hat{v},\hat{v}^{\dagger}$ describe the noise field, i.e.
the filter degrees of freedom.
In this extended space,
the combination $\hat{b}+\hat{v}^{\dagger}$ do commute with
$\hat{b}^{\dagger}+\hat{v}$ and such operators can be measured
simultaneously. In the case of the unimodular phase operator the combination
\begin{equation}
\label{unimodular}
\frac{\hat{b}+\hat{v}^{\dagger} }{\sqrt{
(\hat{b}^{\dagger}+\hat{v})(\hat{b}+\hat{v}^{\dagger})}}
\end{equation} is operationally unique and the extended field modes can be
simultaneously measurable.
A  similar approach to the measurement of a simple quantum system, using
an  extension by an additional degree of freedom, has been used by
Einstein Podolsky and Rosen \cite{epr}. In their case the particle system
with the non commuting position and momentum operators $\hat{p}$ and $\hat
{q}$ has been extended to commuting observables $\hat{p}+\hat{P}$ and
$\hat{q}-\hat{Q}$, where $\hat{P}$ and $\hat{Q}$ can be
seen as the momentum and the position of the filter. This physical procedure
has a counterpart in mathematics, and is called the Naimark extension of the
POVM into a projective measure on a larger Hilbert space \cite{peres}.

The set of operational operators corresponding to the powers
of the operator $\hat{A}$ in the extended filter--system are defined as
\begin{equation}
\hat{A}^{(n)}_{{\cal F}}\equiv
{\rm Tr}_{v}\{\hat{A}^{n}(\hat{b}+\hat{v}^{\dagger},\hat{b}^{\dagger}+\hat{v})
\hat{\varrho}(\hat{v}^{\dagger},\hat{v})\}\,.
\end{equation}
We will assume that the noise field density matrix
$\hat{\varrho}(\hat{v}^{\dagger},\hat{v})$ can be
expressed in the normally order form i.e., all the annihilation operators
are to the right of the creation operators. This is true for all
physically interesting cases discussed in the literature.
Provided that this condition is fulfilled
we can rewrite the above expression, using the coherent state decomposition
of unity, in the following way
\begin{equation}
\label{o_operational}
\hat{A}^{(n)}_{{\cal F}}=
\int\frac{d^{2}\alpha}{\pi}A^{n}(\alpha,\alpha^{\ast})
\hat{\varrho}(\alpha^{\ast}-\hat{b}^{\dagger},\alpha-\hat{b})\,.
\end{equation}
This formula gives us  the form of the POVM for the
experimental NFM  setup with arbitrary noise:
\begin{equation}
\label{filter}
\hat{{\cal
F}}(\alpha)=\hat{\varrho}(\alpha^{\ast}-\hat{b}^{\dagger},\alpha-\hat{b}),
\end{equation}
with the normalization:
\begin{equation}
\int\frac{d^{2}\alpha}{\pi} \hat{{\cal F}}(\alpha) = 1.
\end{equation}
Such a result has a simple meaning in terms of shifts in the coherent state
phase space of the signal field. In order to measure
the signal mode  we have to probe it with a filter state shifted  in the
 phase space by the amount $\alpha$, and sum up over all possible manifolds
of shifts\cite{kw84}.

As an example of this general approach we  find explicitly the POVM for
several interesting cases of the filter state. For the vacuum state
\begin{equation}
\hat{\varrho}(\hat{v}^{\dagger},\hat{v})=|0\rangle\langle 0|=\:
:\exp{(-\hat{v}^{\dagger}\hat{v})}:
\end{equation}
the corresponding POVM is
\begin{equation}
\hat{{\cal F}}=|\alpha\rangle\langle \alpha|,
\end{equation}
where we employed the following identity \cite{louisell}
\begin{equation}
|\alpha\rangle\langle \alpha|=\:
:\exp{[-(\hat{b}^{\dagger}-\alpha^{\ast})(\hat{b}-\alpha)]}:\:,
\end{equation}
and have denoted the normal ordering by  semicolons. This
result has been obtained already in Ref. \cite{bge+kw+pr95}.
Another interesting state is the squeezed vacuum state \cite{sqstates}
\begin{eqnarray}
\nonumber
\hat{\varrho}(\hat{v}^{\dagger},\hat{v})&= &
|s\rangle\langle s|=\cosh^{-1}s \\
&& \times:\exp{\left[-\left(\hat{v}^{\dagger}\hat{v}+
\frac{1}{2}\tanh s\left( e^{i\phi}\hat{v}^{\dagger 2} +
e^{-i\phi}\hat{v}^{2}\right)\right)\right]} :
\end{eqnarray}
which leads to the POVM in the form
\begin{equation}
\hat{{\cal F}}=|\alpha,s\rangle\langle \alpha,s|,
\end{equation}
which has been derived in Ref. \cite{pk+kw96}.
The squeezed state is generated from the ground
state by the action of the squeezing operator $\hat{S}(s,\phi)$
and the Glauber displacement operator $\hat{D}(\alpha)$.
Similarly, for the Fock state $|n\rangle\langle n|$ we obtain
\begin{equation}
\hat{\cal F}=\hat{D}(\alpha)
\frac{1}{\bar{n}+1}
\left(\frac{\bar{n}}{\bar{n}+1}\right)^{\hat{a}^{\dagger}\hat{a}}
\hat{D}^{\dagger}(\alpha),
\end{equation}
which equals to the POVM from the Ref. \cite{bge+kw95} introduced in the
context of the operational position and momentum measurement.

\section{Operational Homodyne Observables and the Position and
Momentum Measurement}
As a first application of the above described formalism we shall
discuss the operational measurement of the homodyne quadratures.
Such a discussion has already been carried in  \cite{kb+kw96}, for
a  homodyne detection in the standard configuration i.e., with one beam
splitter and two detectors. Here we present
the operational observables associated with the joint measurements of two
field quadratures with the help of the NFM apparatus. We will also slightly
generalize
our previous considerations, assuming that the beam splitter, which mixes
the probe field with the noise field has an arbitrary transmittance
coefficient $T$ and that the noise filed is  a squeezed vacuum.
As before,
we shall assume that the local oscillator is a strong coherent filed
described by $\alpha=|\alpha|\exp{(i\theta)}$.

In the extended filter--system space, the  operators leading
to the operational observables of the two homodyne quadratures are
the ``unnormalised'' sine and cosine operators \cite{nfm,pr+kw94}
\begin{equation}
\label{quad1}
\hat{X}_{1}=
\frac{1}{2}\left[
\alpha(\sqrt{T}\hat{b}^{\dagger}+\sqrt{1-T}\hat{v})+
\alpha^{\ast}(\sqrt{T}\hat{b}+\sqrt{1-T}\hat{v}^{\dagger})\right]
\end{equation}
and
\begin{equation}
\label{quad2}
\hat{X}_{2}=
\frac{1}{2}\left[
i\alpha(\sqrt{1-T}\hat{b}^{\dagger}+\sqrt{T}\hat{v})-
i\alpha^{\ast}(\sqrt{1-T}\hat{b}+\sqrt{T}\hat{v}^{\dagger})\right]\,.
\end{equation}
If we define two generating operators ($i=1,2$) for the
corresponding quadratures
\begin{equation}
\hat{Z}_{i\:{\cal F}}(\xi)\equiv
{\rm Tr}_{v}\left\{
\exp{(\xi \hat{X}_{i})}\,|s_{v}\rangle\langle s_{v}|\right\}\,,
\end{equation}
then the operational moments are obtained by the $n$-fold derivatives:
\begin{equation}
\hat{X}^{(n)}_{i\: {\cal F}}=
\left[\frac{d^{n}\hat{Z}_{i\:{\cal F}}(\xi)}{d\xi^{n}}\right]_{\xi=0}\,.
\end{equation}
For both quadratures the generating operators can be easily found
\begin{eqnarray}
\hat{Z}_{1\:{\cal F}}(\xi)&=&
\exp{\left[\frac{\xi\sqrt{T}}{2}\left(
\alpha\hat{b}^{\dagger}+\alpha^{\ast}\hat{b}\right)\right]}
\exp{\left[\frac{\xi^{2}}{8}(1-T)|\alpha|^{2}\sigma_{-}(s)\right]}, \\
\hat{Z}_{2\:{\cal F}}(\xi)&=&
\exp{\left[\frac{\xi\sqrt{1-T}}{2}\left(
i\alpha\hat{b}^{\dagger}-i\alpha^{\ast}\hat{b}\right)\right]}
\exp{\left[\frac{\xi^{2}}{8}(T)|\alpha|^{2}\sigma_{+}(s)\right]},
\end{eqnarray}
where we defined two auxiliary functions
\begin{equation}
\sigma_{\pm}(s)=\cosh 2s \pm \cos(2\theta-\phi)\sinh 2s\
\end{equation}
and denoted by $\phi$ the phase of the squeezed vacuum.
In the first factor of the  generating operators we recognize
the intrinsic quadrature operator
\begin{equation}
\hat{x}_{\theta}=
\frac{e^{i\theta}\hat{b}^{\dagger}+e^{-i\theta}\hat{b}}{\sqrt{2}},
\end{equation}
with the rotated quadrature $\hat{x}_{\theta+\pi/2}$ in $\hat{Z}_{2\:{\cal
F}}(\xi)$.
The second factor of the  generating operators may be seen as an
operator ordering parameter introduced by Cahill and Glauber
\cite{kec+rjg69}. By varying  the filter parameters we can select a
particular ordering, for example for
\begin{equation}
T=\frac{\sigma_{\pm}(s)}
{1+\sigma_{\pm}(s)}
\end{equation}
the creation and the annihilation operators are ordered antinormally
\begin{equation}
\hat{Z}_{1\:{\cal F}}(\xi)=
\exp{((\xi/2)\alpha^{\ast}\sqrt{T}\hat{a})}
\exp{((\xi/2)\alpha\sqrt{T}\hat{a}^{\dagger})}.
\end{equation}
For the discussed experimental NFM
scheme, no selection of the parameters can lead to a normal ordering.

The operational quadratures might be  calculated explicitly
\begin{eqnarray}
\hat{X}^{(n)}_{1\: {\cal F}}&=&
\left(\frac{\sqrt{1-T}}{2i}\sqrt{\sigma_{-}(s)}\right)^{n}
{\rm
H}_{n}\left(i\:\sqrt{\frac{T}{1-T}}\:\frac{\hat{x}_{\theta}}{\sigma_{-}(s)}
\right), \\
\hat{X}^{(n)}_{2\: {\cal F}}&=&
\left(\frac{\sqrt{T}}{2i}\sqrt{\sigma_{+}(s)}\right)^{n}
{\rm H}_{n}\left(i\:\sqrt{\frac{1-T}{T}}\:
\frac{\hat{x}_{\theta+\pi/2}}{\sigma_{+}(s)}\right),
\end{eqnarray}
where ${\rm H}_{n}(z)$ is the $n$-th Hermite polynomial and we have scaled
the  results by setting $\alpha=\sqrt{2}$.
For the purpose of further discussion we write explicitly
the first two operational moments for both quadratures
\begin{eqnarray}
\hat{X}^{(1)}_{1\: {\cal F}}&=&
\sqrt{T}\:\frac{\hat{x}_{\theta}}{\sqrt{\sigma_{-}(s)}}, \\
\hat{X}^{(2)}_{1\: {\cal F}}&=&
T\:\frac{\hat{x}^{2}_{\theta}}{\sigma_{-}(s)}+
\frac{1-T}{2}\,\sigma_{-}(s), \\
\hat{X}^{(1)}_{2\: {\cal F}}&=&
\sqrt{1-T}\:\frac{\hat{x}_{\theta+\pi/2}}{\sqrt{\sigma_{+}(s)}}, \\
\hat{X}^{(2)}_{2\: {\cal F}}&=&
(1-T)\:\frac{\hat{x}^{2}_{\theta+\pi/2}}{\sigma_{+}(s)}+
\frac{T}{2}\,\sigma_{+}(s).
\end{eqnarray}
We see that for $T=1$ or $T=0$ we are restricted to the
measurement of either
$\hat{X}^{(n)}_{1\: {\cal F}}$ or $\hat{X}^{(n)}_{2\: {\cal F}}$,
which are, in this case, equal to the powers of the intrinsic operators.

For simplicity, let assume that the phase of the squeezing vacuum is such
that $2\theta+\phi=0$, then the functions $\sigma_{\pm}(s)$
becomes simpler
\begin{equation}
\sigma_{\pm}(s)=e^{\pm 2s},
\end{equation}
and the operational spread defined as
$\delta\hat{X}_{i\:{\cal F}}\equiv\sqrt{
\langle \hat{X}^{(2)}_{i\: {\cal F}}\rangle-
\langle \hat{X}^{(1)}_{i\: {\cal F}}\rangle^{2}}$
reads for both quadratures
\begin{eqnarray}
(\delta\hat{X}_{1\:{\cal F}})^{2} &=& Te^{2s}(\Delta x_{\theta})^{2}+
(1-T)\:\frac{e^{-2s}}{2}, \\
(\delta\hat{X}_{2\:{\cal F}})^{2} &=& (1-T)e^{-2s}
(\Delta x_{\theta+\pi/2})^{2} + T\:\frac{e^{2s}}{2},
\end{eqnarray}
where $\Delta x_{\theta}$ is the intrinsic quantum dispersion.

We note that for the selected filter parameters one of the
operational quadratures is squeezed and the second one is enhanced. This
behavior is the same as  for the
intrinsic quadratures \cite{sqstates}, however the measuring device
introduces an additional factor contributing to the whole operational spread.
Each operational spread contains two parts, the first part is the
intrinsic uncertainty weighted by the factor $T$ (or $1-T$) and
the second part is the noise  with the weight $1-T$ (or $T$) introduced by
the experimental device.  The origin of these additional terms is clear. The
NFM experimental setup enables us to perform a {\em joint} measurement
of two quadratures in the extended Hilbert space. In the reduced space of the
signal mode we pick up  an additional noise due to the leaking noise mode
through the unused ports in the NFM apparatus.

By  changing the transmittance coefficient
we can select which quadrature will be measured and how precise
this measurement will be.
If we set $T=0$ or $T=1$ we measure only one quadrature and the moments
of the operational operators equal to the powers of the intrinsic homodyne
observables.  The operational result
reduces to the  intrinsic observable in this case because we have assumed
perfect  photodetectors with 100\% efficiency.

The non ideal detectors can be described by a homodyne setup with perfect
detectors, and  a beam splitter with a certain effective transmittivity. The
beam splitter mixes the incoming signal with the vacuum field and the results
of the calculations are formally the same as those carried on for the
non ideal detector \cite{kb+kw96,ul+hp93a}.

It's worth to notice that in fact the NFM experimental
scheme may be  used  for such a description of a  quadrature measurement
with a non ideal photodetector. In such a case we may forget about the
upper right part of the NFM apparatus and  see that the setup reduces to
the situation described in \cite{ul+hp93a}. In such a scheme there is a beam
splitter in front of a pair of ideal detectors. If we put the squeezing
parameter equal to 0, the operational observables
obtained in this paper  reduce to  those obtained in\cite{kb+kw96}.

We see that there are two possible and mathematically equivalent
descriptions of the operational measurement performed with help of the
NFM device.
In one description the experiment is  a
simultaneous measurement of two operational quadratures with the help of
ideal photodetectors with  the NFM apparatus noise.
In the other description, we perform a standard homodyne experiment, with
a non unit quantum efficiency of the photo-detectors simulated by the beam
splitter. Such a beam splitter is mixing the probe signal with the vacuum
state and the effective transmittance coefficient plays the role of the
photodetector efficiency.

Another possible choice of the parameters leads us to the operational
momentum and position operators derived in \cite{bge+kw95}, indeed for
$T=\frac{1}{2}$, $s=0$ and $|\alpha|=2,\,\theta=0$ our results can be reduced
to the special case of the results from this reference, with  the filter
being in the vacuum state.

The of operational operators
$\hat{X}^{(n)}_{1\: {\cal F}}$ become then the operational moments
of the position operator whereas $\hat{X}^{(n)}_{2\: {\cal F}}$
the operational moments of the momentum operator.

\section{Squeezed Quantum Trigonometry}
Originally the NFM experiments have been devoted to the measurement of
the relative phase between a high--intensity classical field and a
low--intensity laser field. The quantities which were investigated
in these experiments were built out of the field quadratures (\ref{quad1})
and (\ref{quad2}) ( with transmittance $T=\frac{1}{2}$).
These operators have been  constructed in such a way that they  can play
the role of the commuting ``sine'' and ``cosine'' operators.
As a result the phase of the quantum probe field has been defined
operationally, avoiding all the problems associated with the construction of
a Hermitian phase operator in quantum mechanics\cite{lynch95}.
It was shown \cite{bge+kw+pr95} that one can associate
with the NFM experiment an operational phase operator.

In this section we extend this  theoretical analysis for the case
of the noise field being in a squeezed vacuum state (the original NFM
experiments have been done with the vacuum state). We will show how to define
an operator corresponding
to any function of the phase. As a example we derive operators corresponding
to the trigonometric functions. These operators form the squeezed quantum
trigonometry of operational observables for the NMF setup.

We shall start our discussion, introducing an amplitude marginal of the POVM
for the NFM device
\begin{equation}
\hat{{\cal F}} (\varphi) =\int dI \hat{{\cal F}}(\alpha=
\sqrt{I}e^{i\varphi}).
\end{equation}

From this POVM, we construct the corresponding phase propensity
\begin{equation}
\Pr(\varphi)=\langle \hat{{\cal F}}(\varphi)\rangle,
\end{equation}
which is a periodic function of phase, normalized to unity
\begin{equation}
\label{normalization}
\int \frac{d\varphi}{2\pi}\Pr(\varphi)=1.
\end{equation}

\subsection{Phasor Basis}
We define here the set of operational operators, which in the classical limit
can be reduced to the Fourier expansion basis.
The problem of constructing the
 operators corresponding to the classical phase dependent quantities has
been solved in Ref. \cite{jb+bge91}. Because every periodic function can be
expressed in terms of its Fourier decomposition
\begin{equation}
f(\varphi)=\sum_{k=-\infty}^{\infty}f_{k}e^{ik\varphi}\,,
\end{equation}
we will construct the operational quantum analogs of
$e^{ik\varphi}$ --- so called {\em phasors} --- in the following way
\begin{equation}
\label{meanv-phasor-def}
\overline{e^{ik\varphi}}\equiv\langle \hat{E}^{(k)}\rangle,
\;\;\; k=\pm1, \pm2, \ldots.
\end{equation}
The reality of the propensity implies
\begin{equation}
\label{reality}
\hat{E}^{(-k)}=\hat{E}^{(k)\,\dagger},
\end{equation}
because of the normalization condition (\ref{normalization}) we have
$\hat{E}^{(0)}=1$.

As we have seen in the previous sections, the function $f(\varphi)$
corresponds in the NFM device to the following operational operator
\begin{equation}
\label{fourier}
\hat F=\sum_{k=-\infty}^{\infty}f_{k}\hat{E}^{(k)},
\end{equation}
which is a function of the boson creation and annihilation operators of the
probe field.

Because we have derived the POVM operator $\hat{{\cal F}}$ for
the NFM setup with an arbitrary noise field (\ref{filter}), we can
derive the explicit expression for the phasor
$\hat{E}^{(k)}$ straight from the definition (\ref{op_operator}).
Once these operators are known, the operational quantum operator can be
calculated using a Fourier series expansion.
Because of this we will call the set $\{\hat{E}^{(k)},\,k=0,1,2,\ldots\}$
the phasor basis. The definition of the phasors introduced in \cite{jb+bge91}
requires, that these operators become in the classical limit  functions of
the phase only.

In order to find the phasor basis we start with the formula
obtained in \cite{bge+kw+pr95}
\begin{equation}
\label{phasor-def}
E^{(k)}=\left(\frac{\alpha}{\alpha^*}\right)^{k/2}
{\rm Tr}_v\left\{U^k\,\hat{S}(s,\phi)|0_{v}\rangle\langle 0_{v}|
\hat{S}^{\dagger}(s,\phi)\right\},
\end{equation}
where we have replaced  the vacuum state of the filter by the
squeezed vacuum state.

The complex number $\alpha$ denotes the amplitude of the classical
reference field of the local oscillator.  The unitary operator
$\hat{U}$ is defined as follows
\begin{equation}
\label{U-def}
\hat{U}=\frac{\hat{X}_{1}+i\hat{X}_{2}}{\sqrt{\hat{X}_{1}^2+\hat{X}_{2}^2}}=
\frac{\sqrt{\hat{X}_{1}^2+\hat{X}_{2}^2}}{\hat{X}_{1}-i\hat{X}_{2}},
\end{equation}
where $\hat{X}_{i}$ are given by (\ref{quad1}) and (\ref{quad2}) with
the beam splitter transmittivity $T=\frac{1}{2}$. Simple algebra gives
\begin{equation}
\hat{U}=\sqrt{\alpha^*/\alpha}
\frac{\hat{b}+\hat{v}^{\dagger}}
{\sqrt{(\hat{b}+\hat{v}^{\dagger})(\hat{b}^{\dagger}+\hat{v})}}
=\sqrt{\alpha^*/\alpha}\left(\frac{\hat{b}+\hat{v}^{\dagger}}
{\hat{b}^{\dagger}+\hat{v}}\right)^{1/2}\,.
\end{equation}
We recognize in this expression the unimodular operator (\ref{unimodular}).
In this way we obtain an equivalent
expression for $\hat{E}^{(k)}$
\begin{equation}
\label{phasor}
\hat{E}^{(k)}(s,\phi)=
{\rm Tr}_{v}\left\{\frac{(\hat{b}+\hat{v}^{\dagger})^{k}}
{\left[(\hat{b}+\hat{v}^{\dagger})
(\hat{b}^{\dagger}+\hat{v})\right]^{\frac{k}{2}}}
\hat{S}(s,\phi)|0\rangle\langle 0|\hat{S}^{\dagger}(s,\phi)
\right\},
\end{equation}
with $k=1,2,3, \ldots$. Negative values of $k$
in (\ref{phasor}) may be derived using (\ref{reality}).

Because operator $\hat{U}$ obeys the assumptions imposed on
the operator  $\hat{A}(\hat{b}+\hat{v}^{\dagger},\hat{b}^{\dagger}+\hat{v})$
defined in (\ref{o}) we can write the formula for the phasor using
(\ref{o_operational}) as follows:
\begin{equation}
\label{phasor-int}
\hat{E}^{(k)}(s,\phi)=
\int\frac{d^{2}\alpha}{\pi}\frac{\alpha^{k}}{(\alpha^{\ast}\alpha)^{\frac{k}{2}}}
|\alpha,s\rangle\langle \alpha,s|\,,
\end{equation}
obviously $\hat{{\cal F}}(s,\phi)=|\alpha,s\rangle\langle \alpha,s|$
is the POVM associated with the described experimental scheme.
It is clear, that the POVM, contrary to spectral measure, depends on
experimental device. In fact,
for each value of squeezing parameter $s$ we obtain a
different propensity a different POVM and a different phasor basis,
even though the probe field remains unchanged.

An exact formula for the phasor may be derived straight
from (\ref{phasor-int}),
\begin{equation}
\label{phasor-formula}
\hat{E}^{(k)}(s,\phi)=\hat{S}(s,-\phi)\vdots
\left[\frac{(\hat{b}\cosh s-\hat{b}^{\dagger}e^{i\phi}\sinh s)^{k}}
{\hat{b}^{\dagger}\cosh s-\hat{b}e^{-i\phi}\sinh s)}\right]^{\frac{k}{2}}
\vdots \hat{S}^{\dagger}(s,-\phi).
\end{equation}
where $\vdots$ denotes the antinormal ordering. When the squeezing
parameter $s$ tends to zero the above formula reduces to the  one
from \cite{bge+kw+pr95}.

\subsection{Properties of the Squeezed Quantum Trigonometry }
In this section we find various phase space representations for the
simple combinations  of the phasor basis. Using these relations we derive the
 trigonometric operational operators.

These trigonometric operators will be defined in accordance with
(\ref{fourier}). We have the following operators
\begin{eqnarray}
\label{trigfun-def}
\hat{S}^{(1)} \equiv \frac{1}{2i}(\hat{E}^{(1)}-\hat{E}^{(-1)}) \,,\,\,\,\,
\hat{S}^{(2)}\equiv \frac{1}{2}-
\frac{1}{4}(\hat{E}^{(2)}+\hat{E}^{(-2)})\,, \nonumber \\
\hat{C}^{(1)}\equiv \frac{1}{2}(\hat{E}^{(1)}+\hat{E}^{(-1)}) \,,\,\,\,\,
\hat{C}^{(2)}\equiv \frac{1}{2}+\frac{1}{4}(\hat{E}^{(2)}+\hat{E}^{(-2)})\,,
\end{eqnarray}
that correspond to the trigonometric functions
$\sin{\varphi},\,\cos{\varphi},\,\sin^{2}{\varphi}$ and
$\cos^{2}{\varphi}$ respectively.
As it was stressed earlier and
follows from definition of the phasors $(\hat{C}^{(1)})^{2}\neq\hat{C}^{(2)}$.
Consequently, the Pythagorean theorem
$\sin^{2}{\varphi}+\cos^{2}{\varphi}=1$ is obeyed by
$\hat{S}^{(2)},\,\hat{C}^{(2)}$, but not by
$(\hat{S}^{(1)})^2$ and $(\hat{C}^{(1)})^2)$.

In order to investigate the properties of the phasors we
evaluate their phase space representations
corresponding to a certain arbitrary $\eta$-ordering of boson operators,
such a phase representation may be defined as follows \cite{kec+rjg69}
\begin{equation}
\label{phasorP}
E^{(k)}_{P^{(\eta)}}(s,\phi)=\int\frac{d^{2}\beta}{\pi}
e^{\alpha\beta^{\ast}-\alpha^{\ast}\beta}e^{\frac{\eta}{2}|\beta|^{2}}
{\rm Tr}\left\{\hat{E}^{(k)}(s,\phi)\hat{D}(\beta)\right\}.
\end{equation}
The number $\eta$ lies between minus infinity and plus one and $P^{(\eta)}$
denotes  the corresponding quasidistribution. Especially
interesting cases are $P^{(-1)},P^{(0)},P^{(-1)}$, which correspond to the
$Q$-representation (normal ordering), the Wigner function (symmetric
ordering) and the $P$-representation (antinormal ordering), respectively.
In this formula the phase space integration can be done and as a result we
obtain
\begin{eqnarray}
\nonumber
E^{(k)}_{P^{(\eta)}}(s,\phi)&=&
\int_{-\infty}^{\infty}\frac{d^{k}\lambda}{\sqrt{\pi^{k}}}
\Omega^{\frac{k+1}{2}}{\rm H}_{k}\left(
\frac{1}{2}\sqrt{\Omega}\left[\Sigma\beta+\lambda^{2}\beta^{\ast}
e^{-i\phi}\sinh 2s\right]\right) \\
\label{phasorp2}
& & \times\exp{\left\{-\lambda^{2}\Omega\left[\Sigma|\beta|^{2}+
\frac{1}{4}\lambda^{2}\sinh 2s\left(e^{i\phi}\beta^{2}+e^{-i\phi}
\beta^{\ast \:2}\right)\right]\right\}}\,,
\end{eqnarray}
where
\begin{equation}
\Omega\equiv
\left[\left(e^{-s}\left(1-\lambda^{2}\frac{\eta+1}{2}\right)
+\lambda^{2}\cosh s\right)
\left(e^{s}\left(1-\lambda^{2}\frac{\eta+1}{2}\right)
+\lambda^{2}\cosh s\right)\right]^{-1}
\end{equation}
and
\begin{equation}
\Sigma\equiv\lambda^{2}\left(\cosh^{2}s-\frac{\eta+1}{2}\right)+1\,.
\end{equation}

This result is meaningful only if $\Omega$ is real, otherwise one
of the integrals, that appeared earlier, is divergent.
Because the most interesting cases are the $Q,\,P$-representations and the
Wigner function we have
\begin{equation}
\Omega=\left\{
\begin{array}{ccl}
\left[\left(\lambda^{4}+2\lambda^{2}\right)\cosh^{2}s+1\right]^{-1} &
{\rm for} & \eta=-1 \\
\left[\frac{1}{4}\lambda^{4}+\lambda^{2}(2\sinh^{2}s+1)+1\right]^{-1} &
{\rm for} & \eta=0 \\
\left[1-\left(\lambda^{4}+2\lambda^{2}\right)\sinh^{2}s\right]^{-1} &
{\rm for} & \eta=1\,.
\end{array}
\right.
\end{equation}
For $\eta=1$, $\Omega$ becomes a complex number, so the
$P$-representation for the squeezed phasors does not exists unless we tend
with $s$ to zero. In this limit we retrieve the original NFM phasors from
\cite{bge+kw+pr95}.
In our further analysis we shall restrict the calculations to the Wigner
function only, but all the results may be easily rewritten for the respective
$Q$-representation.

If we write $\beta=\sqrt{I}e^{i\theta}$ and use (\ref{phasorp2}),
the definitions (\ref{trigfun-def}) give for $\eta=0$
the ``sine'' and the ``cosine'' Wigner functions
\begin{eqnarray}
C^{(1)}_{W}(s,\phi)&=&\int_{-\infty}^{\infty}\frac{d\lambda}{\sqrt{\pi}}
\exp{\left\{\frac{-\lambda^{2}I[1-\frac{1}{2}\lambda^{2}+
\lambda^{2}\cosh^{2}s(1+\tanh s \cos(\phi+2\theta))]}
{\frac{1}{4}\lambda^{4}+\lambda^{2}(1+2\sinh^{2}s)+1}\right\}} \nonumber \\
\label{cos1w}
& &\times \frac{\sqrt{I}\left[\cos\theta-\frac{1}{2}\lambda^{2}\cos\theta
+\lambda^{2}\cosh^{2}s(\cos\theta+\tanh s \cos(\phi+\theta))\right]}
{[\frac{1}{4}\lambda^{4}+\lambda^{2}(1+2\sinh^{2}s)+1 ]^{\frac{3}{2}}}
\end{eqnarray}
and
\begin{eqnarray}
S^{(1)}_{W}(s,\phi)&=&\int_{-\infty}^{\infty}\frac{d\lambda}{\sqrt{\pi}}
\exp{\left\{\frac{-\lambda^{2}I[1-\frac{1}{2}\lambda^{2}+
\lambda^{2}\cosh^{2}s(1+\tanh s \cos(\phi+2\theta))]}
{\frac{1}{4}\lambda^{4}+\lambda^{2}(1+2\sinh^{2}s)+1}\right\}} \nonumber \\
\label{sin1w}
& &\times\frac{\sqrt{I}\left[\sin\theta-\frac{1}{2}\lambda^{2}\sin\theta+
\lambda^{2}\cosh^{2}s(\sin\theta-\tanh s \sin(\phi+\theta))\right]}
{[\frac{1}{4}\lambda^{4}+\lambda^{2}(1+2\sinh^{2}s)+1 ]^{\frac{3}{2}}}.
\end{eqnarray}
Similar expressions for the Wigner functions $S^{(2)}_{W}(s,\phi)$ and
$C^{(2)}_{W}(s,\phi)$ can be derived from the phasor basis. Due to the
lengthy mathematical form, we shall present in this paper only plots of
these functions.
The Wigner functions of the operational quantum trigonometry provide a simple
framework for the physical discussion of the obtained results.
In Fig. 2, we have plotted the Wigner function
$S^{(1)}_{W}(2,\frac{\pi}{2})$.
For field intensities tending to infinity we should reproduce the  classical
trigonometry. In order to see this we recall that $\lim_{I\rightarrow\infty}
\left(\sqrt{\frac{I}{\pi}}\right)e^{-I f^{2}(z)}=\delta(f(z))$.
After some simple  algebra
we see that  $f(\lambda)$ in (\ref{cos1w}) and (\ref{sin1w}) has only
one zero point at $\lambda_{0}=0$ and the modulus of a derivative of
$f(\lambda)$ in that point is one. As a result, using the well known relation
between $\delta(f(z))$ and $\delta(z)$, we see that in order to perform the
classical limit, we have to  put only $\lambda=0$ in the integrands of the
Wigner functions. In this limit
\begin{eqnarray}
\nonumber
\lim_{I\rightarrow\infty}S^{(1)}_{Q\, {\rm or}\, W}(s,\phi)&=&\sin\theta \\
\label{infinity-asypt}
\lim_{I\rightarrow\infty}C^{(1)}_{Q\, {\rm or}\, W}(s,\phi)&=&\cos\theta.
\end{eqnarray}
This result is also seen on Fig 2. where indeed for
large values $I$  we reconstruct the classical  trigonometric function.

As it might have been expected, in the classical limit the dependence on
the squeezing parameters disappears, in agreement with the fact that
squeezing is a purely quantum effect.

Similarly we can prove, that
$\hat{S}^{(2)}(s,\phi)$ and $\hat{C}^{(2)}(s,\phi)$ posses a proper
classical limit. For these functions we present only graphs. As it can be
seen  on Fig. 3 and Fig. 4,  the classical limit of the Wigner function of
$\hat{C}^{(2)}(s,\phi)$ reproduces the $\cos^{2}\varphi$.

We  shall discuss now the  limit of a very small probe
field intensity. Let assume for a while, that the phase of squeezing
parameter $\phi$ is zero, and  ${I\rightarrow 0}$. In this case we can
expand the exponential factor of the integrand in (\ref{sin1w}) and
(\ref{cos1w}). Keeping only the first term  we obtain
\begin{eqnarray}
\nonumber
C^{(1)}_{W}(s,\phi)&\stackrel{I\rightarrow 0}{=}&
\sqrt{I}{\cal A}_{+}(s,\phi=0)\cos\theta \\
S^{(1)}_{W}(s,\phi)&\stackrel{I\rightarrow 0}{=}&
\sqrt{I}{\cal A}_{-}(s,\phi=0)\sin\theta
\end{eqnarray}
where
\begin{equation}
\label{amplitude}
{\cal A}_{\pm}(s,\phi=0)=
\int_{-\infty}^{\infty}\frac{d\lambda}{\sqrt{\pi}}
\frac{1+(1/2)\lambda^{2}e^{\pm 2s}}
{[\frac{1}{4}\lambda^{4}+\lambda^{2}(1+2\sinh^{2}s)+1 ]^{\frac{3}{2}}}
\end{equation}
are two  different amplitudes for the ``sine'' and the ``cosine'' Wigner
functions. These
amplitudes depend only on the squeezing parameter. This result
shows the effect of the amplitude squeezing near the vacuum. From
(\ref{amplitude}) we see that for $\phi=0$ $S^{(1)}_{W}(s,0)$ is strongly
squeezed then $C^{(1)}_{W}(s,0)$ (i.e. the amplitude of sine is smaller then
that for cosine). Obviously for $I=0$ the  amplitudes are  exactly equal to
zero, which is in agreement with the common intuition that
the phase of a light field in the vacuum state is randomly distributed.

Changing  the squeezing phase to $\phi=\pi$, we reverse the amplitude
squeezing from the sine to the cosine function. It is simple to verify,
that the amplitudes for
different values of $\phi$ are connected in the following way
${\cal A}_{+}(s,\phi=0)={\cal A}_{-}(s,\phi=\pi)$
and ${\cal A}_{-}(s,\phi=0)={\cal A}_{+}(s,\phi=\pi)$, we see that in this
case, that $C^{(1)}_{W}(s,0)$ is strongly squeezed then $S^{(1)}_{W}(s,0)$.

For arbitrary values of $\phi$ and $s$ the separation of the cosine
Wigner function into an amplitude, and a purely angle dependent part
is no longer possible, but a clear squeezing of the amplitude can also
observed comparing, for example, Fig. 3 (s=0.5) and Fig. 4 (s=1.5).
For $s=2$ (Fig. 2) the operational phasor is by far more squeezed then the
two functions form Fig. 3 and Fig. 4. Note that  the range of the
intensity $I$ is almost twice bigger then the range for Fig. 3 and 4.

Similarly we can find an asymptotic expression for $C^{(2)}_{W}(s,\phi)$.
In the limit of small $I$
\begin{equation}
C^{(2)}_{W}(s,\phi)=\frac{1}{2}(1-c(s,\phi)),
\end{equation}
where $c(s,\phi)$ is an $I$-independent function of the squeezing parameter
\begin{equation}
c(s,\phi)=\int_{-\infty}^{\infty}\frac{d\lambda_{1}d\lambda_{2}}{\pi}
\frac{(1/2)(1+(1/4)\lambda^{4}+\lambda^{2}(1+2\sinh^{2}s))\sinh 2s\cos\phi}
{[\frac{1}{4}\lambda^{4}+\lambda^{2}(1+2\sinh^{2}s)+1]^{\frac{5}{2}}},
\end{equation}
where $\lambda$ is understood here as a norm of the two component vector
$\vec{\lambda}=[\lambda_{1},\lambda_{2}]$.
It's easy to check, that $c(s,\phi)$ tends either to unity ($\phi=0$)
or to minus unity ($\phi=\pi$).
As a result, for small intensities $C^{(2)}_{W}(s,\phi)$
becomes zero ($\phi=0$) or one ($\phi=\pi$).
For $\phi=\pi/2,\,3\pi/2$ we have $c(s,\phi)=0$ and
$\lim_{I\rightarrow 0}C^{(2)}_{W}(s,\phi)=1/2$.

For small $I$ the squeezing
influences the system very strongly. If the squeezed phase $\phi$ equals
to $\pi/2$, the $\hat{C}^{(2)}$ Wigner
function tends to $1/2$
(Fig. 3), whereas for $\phi=0$ it takes values near  zero
(Fig. 4). Such a dramatic change of the cosine quadratures occurs because
in the limit of small $I$,  purely quantum effects of the squeezed vacuum are
important.
The squeezing allows one of the quadratures to be reduced below
the vacuum level represented by a uniformly distributed random phase.
The uniform distribution of the phase corresponding to the vacuum state
leads to an operational Wigner function for $I=0$ equal to $\frac{1}{2}$.
For a squeezed vacuum, this uniform random-phase distribution is modified
\cite{db+kw92} and a significant change of the operational quadrature is
possible.
In fact fluctuations below $\frac{1}{2}$ in the $\hat{C}^{(2)}$ Wigner
function exhibit the quantum nature of the squeezed vacuum.
Those results are even more clear in view of our previous analysis
of operational quadratures (\ref{quad1}) and (\ref{quad2}). We see
straight from the definition of $\hat{U}$ (\ref{U-def}), that
operational operators $\hat{C}^{(2)}(s,\phi)$ and $\hat{S}^{(2)}(s,\phi)$
are just normalized homodyne quadratures, so their dependence
on squeezing phase and $s$ may be understood as if they were
homodyne quadratures. These operators are also quantum analogs of certain
classical functions of the phase of the electromagnetic field. Because of
this they posses a much richer structure and this enable, for example, to
investigate the phase properties of the quantum optical fields.

\section{Conclusions}
We have presented an operational theory of the eight-port homodyne detection
scheme. For such a measuring apparatus, we have derived the operational
quadrature operators for an arbitrary noise field leaking through the unused
port. In the framework of the operational approach we have derived the
quantum propensity, the  POVM and the corresponding operational observables
for the NFM device. For the noise field being the squeezed vacuum we have
constructed the quantum trigonometry and the corresponding Wigner functions.
\section*{Acknowledgments}
This work has been partially supported
by the Polish KBN grant 2 PO3B 006 11.

\begin{figure}
\begin{center}
\begin{tabular}{rcl}
\leavevmode  \epsfxsize=9.0cm \epsffile{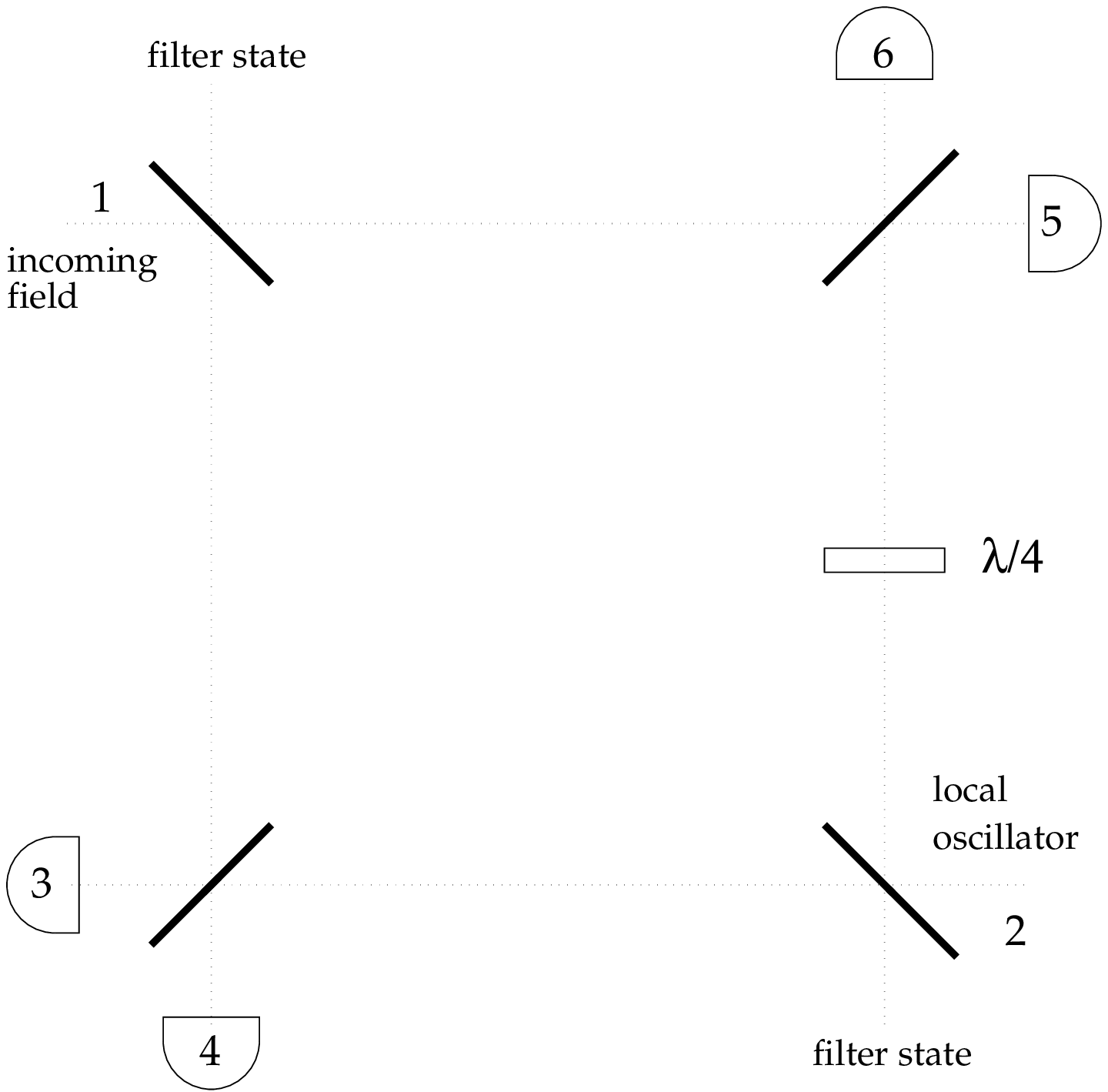} \\
\end{tabular}
\end{center}
\caption{The NFM apparatus. The photodetectors are labeled by numbers $3,4$
and $5,6$, whereas the beam splitters are represented by dark thickened lines}
\label{nfm}
\end{figure}

\begin{figure}[t]
\begin{center}
\begin{tabular}{rcl}
\put(84,45){$I$}
\put(225,80){$\varphi$}
\leavevmode  \epsfxsize=9.0cm \epsffile{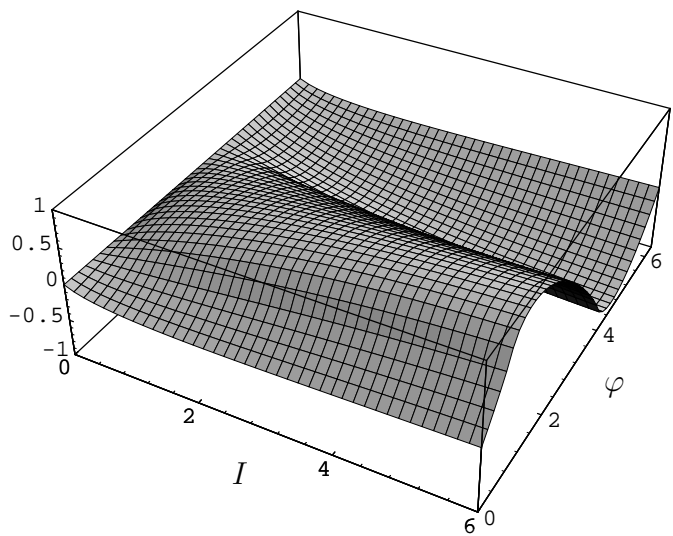} \\
\end{tabular}
\end{center}
\caption{Plot of the Wigner function of the operational operator
$\hat{S}^{(1)}(s=2,\phi=\pi/2)$.}
\label{sin2}
\end{figure}

\begin{figure}[t]
\begin{center}
\begin{tabular}{rcl}
\put(84,45){$I$}
\put(225,80){$\varphi$}
\leavevmode  \epsfxsize=9.0cm \epsffile{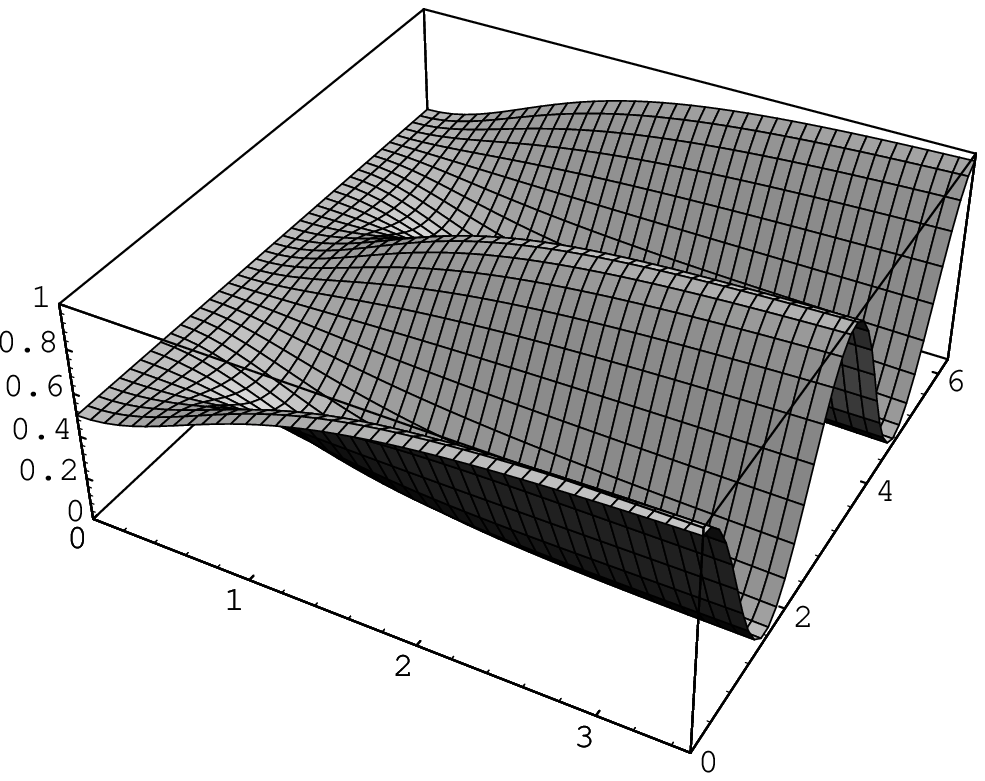} \\
\end{tabular}
\end{center}
\caption{Plot of the Wigner function of the operational operator
$\hat{C}^{(2)}(s=0.5,\phi=\pi/2)$.}
\label{cos2a}
\end{figure}

\begin{figure}[t]
\begin{center}
\begin{tabular}{rcl}
\put(84,45){$I$}
\put(225,80){$\varphi$}
\leavevmode  \epsfxsize=9.0cm \epsffile{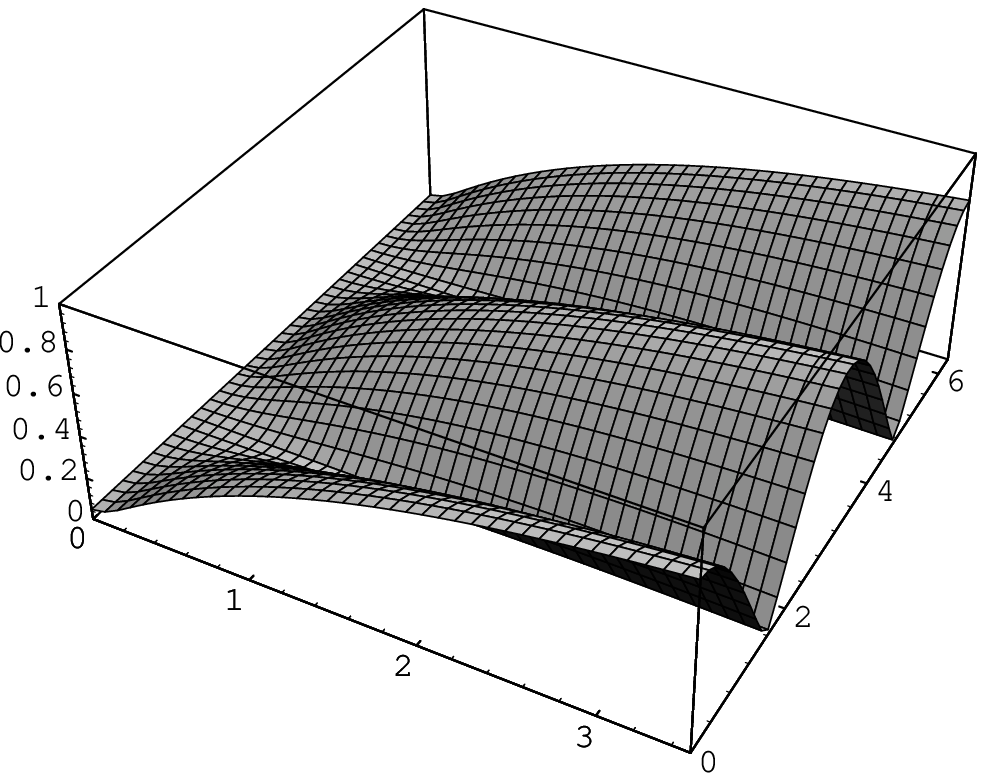} \\
\end{tabular}
\end{center}
\caption{Plot of the Wigner function of the operational operator
$\hat{C}^{(2)}(s=1.5,\phi=0)$.}
\label{cos2b}
\end{figure}

\end{document}